\documentstyle[12pt]{article}

\newcommand{\beq}{\begin{equation}}
\newcommand{\eeq}{\end{equation}}

\begin{document}

\title{Quantum Reed-Muller Codes}
\author{A. M. Steane\\
Department of Physics, University of Oxford\\
Clarendon Laboratory, Parks Road, Oxford OX1 3PU, England.}
\date{21 August 1996}
\maketitle

\begin{abstract}
A set of quantum error correcting codes based on classical Reed-Muller
codes is described. The codes have parameters $[[n,k,d]] =
[[2^r,\; 2^r - C(r,t) - 2\sum_{i=0}^{t-1} C(r,i), \;
2^t + 2^{t-1} ]]$.
  \end{abstract} 

The study of quantum information is currently stimulating much interest.
Most of the basic concepts of classical information theory have
counterparts in quantum information theory, and among these
is the idea of an error correcting code. An error correcting code
is a means of storing information (whether quantum or classical) in
a set of bits (ie either qubits or classical bits) in such a way that the 
information can be extracted even after a subset of the bits has changed
in an unknown way. Such codes are a fundamental part of the study of classical
information channels. The possibility of quantum error correction was only
recently discovered \cite{Shor1,Steane1}. Importantly, it was shown
that efficient quantum codes exist for arbitrarily large amounts of quantum 
information \cite{Cald1,Steane2}. The word `efficient' refers to the fact that 
the rate $k/n$ of the code need not fall off as $n$ increases, for a given 
ratio $d/n$, where $d$ is the minimum distance of the code. This
and other features makes quantum error correction the best
prospect for enabling quantum information to 
be transmitted or stored with a small amount of error, and consequently the 
best prospect for controlling noise in a quantum information processor.

The subject of quantum error correction may be considered to have two distinct 
parts. The first part is to show how to apply error correction in 
a physical situation, and the second is to find good quantum error correcting
codes. This paper is concerned with the second part, that of finding codes.

Following \cite{CaldGF4}, I will use the notation $[[n,k,d]]$ to refer
to a quantum error correcting code for $n$ qubits having $2^k$ codewords
and minimum distance $d$ (previously I used $\{n,k,d\}$ \cite{SQC}).
Such a code enables the quantum information
to be restored after any set of up to $\lfloor (d-1)/2 \rfloor$ qubits
has undergone errors. In addition, when $d$ is even, $d/2$ errors can be
detected. Attention will be limited to the `worst case' that any
defecting qubit (ie any qubit undergoing an unknown interaction)
might change state in a completely unknown way, so all the error
processes $X$, $Z$ and $Y=XZ$ must be correctable
\cite{Steane2,Ekert96,Knill,Gott}, where
  \beq
X = \left( \begin{array}{cc} 0&1\\1&0 \end{array} \right), \;
Y = \left( \begin{array}{cr} 0&-1\\1&0 \end{array} \right), \;
Z = \left( \begin{array}{cr} 1&0\\0&-1 \end{array} \right).
  \eeq

The first known general code construction used a pair of classical codes
$C_1 = [n,k_1,d]$, $C_2 = [n,k_2,d]$ with $C_2^{\perp} < C_1$ to produce a quantum $[[n,k_1 + k_2 - n,d]]$ code\footnote{The `less than' sign
is used here to mean that one linear vector space is a
subspace of the other.} \cite{Cald1,Steane2}. Further
individual quantum codes were subsequently
found \cite{Laf,Benn,Cald2,SQC} which are more efficient than those
given by this construction. Gottesman \cite{Gott} found an infinite set
of optimal single-error correcting quantum codes, with parameters
$[[2^r, 2^r - r - 2, 3]]$. In this paper I will present a set of
quantum codes of which Gottesman's is a subset. They are obtained
by combining classical Reed-Muller codes \cite{MacW}, and have parameters
  \begin{equation}
\left[ \left[ n,\; k, \; d \right] \right] = 
\left[ \left[ 2^r  \rule{0em}{1em}, \; \right. \right.
2^r - C(r,t) - 2\sum_{i=0}^{t-1} C(r,i), \;
\left. \left. 2^t + 2^{t-1} \rule{0em}{1em} \right] \right],  \label{par}
  \end{equation}
where $C(r,t) = r!/t!(r-t)!$. 
These parameters are tabulated in table 1 for small values of $r$ and $t$.

To show how the codes are derived, we will use the results and notation
of Calderbank {\em et al} \cite{Cald2} who showed how to reduce the
quantum coding problem to one of orthogonal geometry. A quantum code
for $n$ qubits
is specified by its generator matrix which has the general form
  \beq
{\cal G} = \left( \left. G_x \right| G_z \right)   \label{Ggen}
  \eeq
where $G_x$ and $G_z$ generate $n$-bit binary vector spaces. The rows of $G_x$ 
and $G_z$ have length $n$, and the number of rows is $n+k$. The minimum 
distance of the code is the minimum weight of a non-zero generated code word, 
where the weight is the number of non-zero bit locations. A bit location is 
non-zero if it is non-zero in either the left hand ($X$) part or the right hand 
($Z$) part. In other words, if a code word is written $( u_x | u_z )$, where 
$u_x$ and $u_z$ are $n$-bit strings, then the weight is the Hamming weight of 
the bitwise {\sc or} of $u_x$ with $u_z$.

To qualify as an error correcting code, the quantum code 
must satisfy a property
which is best specified in terms its {\em stabiliser} 
  \beq
{\cal H} = \left( \left. H_x \right| H_z \right).
  \eeq 
The stabiliser was discussed in \cite{Gott}, it is related to the
generator by \cite{Cald2}
   \beq
H_x \cdot G_z^T + H_z \cdot G_x^T = {\bf 0},   \label{HG}
   \eeq
where the arithmatic is over a binary field (ie $\cdot$ is bitwise {\sc and}
and $+$ is bitwise {\sc xor}). From this relation, it is clear that
the stabiliser and generator are the quantities corresponding 
to the parity check and generator matrices for a classical code
(relation (\ref{HG}) states that ${\cal H}$ may be obtained from
${\cal G}$ by swapping the $X$ and $Z$ parts, and extracting the dual
of the resulting $(n+k) \times 2n$ binary matrix). The
further property which a quantum code must satisfy is \cite{Cald2}
  \beq
H_x \cdot H_z^T + H_z \cdot H_x^T = {\bf 0}.  \label{selfdual}
  \eeq

The encoding method of \cite{Cald1,Steane2} using pairs of classical codes 
leads to a generator and stabilisor of the form
  \beq
{\cal G} = \left( \begin{array}{c|c} G_1 & 0 \\ 0 & G_2 \end{array}
\right), \;\;\;\;\;
{\cal H} = \left( \begin{array}{c|c} H_2 & 0 \\ 0 & H_1 \end{array}
\right)
    \label{G1G2}
  \eeq
where $G_1$ and $G_2$ generate the classical codes $C_1$ and $C_2$.
It is seen that the relation (\ref{selfdual}) is the generalisation
of the `dual' condition $C_2^{\perp} < C_1$ which such codes must satisfy.
Previously I suggested extending these codes by adding further rows
$D$ to $G_1$ such that $G_1$ and $D$ together generate a classical code
of smaller minimum distance than $G_1$ alone, and then introducing sign
changes among the code vectors in a systematic way \cite{SQC}. This method 
corresponds to a generator of the form 
  \beq
{\cal G} = \left( \begin{array}{c|c} G_1 & 0 \\ 0 & G_2 \\ D_x & D_z
           \end{array} \right)   \label{GD}
  \eeq
where the matrix $D$ of \cite{SQC} becomes $D_x$ here. This method
is adopted in what follows.

The following procedure will generate a quantum code having
parameters given in equation (\ref{par}). Explicit examples are
given after the general construction. In equation (\ref{GD}),
take 
  \beq
G_1 = G_2 = \left[ 2^r,\;k_{\rm RM}(t,r),\; 2^{t+1} \right] 
\;\;\; \mbox{Reed-Muller code},
  \eeq
where
  \beq
k_{\rm RM}(t,r) = 2^r - \sum_{i=0}^{t} C(r,i).
  \eeq
Take $D_x$ such that $G_1$ and $D_x$ together generate the Reed-Muller code of 
distance $2^t$. This classical code has size $k_{\rm RM}(t-1,r)$
so the total number of rows in the quantum generator (equation (\ref{GD}))
is
  \beq
n+k = k_{\rm RM}(t,r) + k_{\rm RM}(t-1,r)
  \eeq
which gives the size $k$ of the quantum code in equation (\ref{par}).
If $D_z$ were zero, then the generated quantum code would have
minimum distance $2^t$. We now construct $D_z$ so as to increase this
minimum distance to $2^t + 2^{t-1}$. Let $D_x^i$ refer to the
$i$'th row of $D_x$, and let the rows be numbered $1$ to $m$
where $m = k_{\rm RM}(t-1,r) - k_{\rm RM}(t,r)$. 
Form $D_z$ such that
  \beq
\begin{array}{rcll}
D_z^i &=& D_x^{i+1} \;\;\;           &\mbox{for} \;\;1 \le i < m   \\
D_z^m &=& L_t \left(  D_x^1  \right) &
\end{array}
  \label{Dz}   \eeq
where the action of $L_t$ is to rotate the bit string left by $t$ places
(eg $L_2(00010011) = 01001100$). This completes the construction of
the generator ${\cal G}$, which completely specifies the quantum code.

The proof that these codes have minimum distance $2^t + 2^{t-1}$ for
all $r$ is by induction on $t$, using the fact that this was proved
for $t=1$ previously. The induction on $t$ (ie the fact that $2^t + 2^{t-1}$
is the correct formula for $d$) is obvious from the construction
of the codes. An alternative proof should be possible by induction on 
$r$, following a similar method to a proof for classical Reed-Muller
codes \cite{MacW}, though I have not yet established this explicitly.
To prove that the codes satisfy the `self dual' condition (\ref{selfdual}),
notice that the stabiliser ${\cal H}$ of a code $[[n,k,d]]$ constructed
as above is equal to the generator of a $[[n,-k,d']]$ code constructed
by the same method
(where $d' = 2 d$, ($4 d$) for odd, (even) $r$ respectively,
see table 1). In other words, 
the stabiliser matrix is also built out of classical Reed-Muller
matrices by the same recipe. Because the classical Reed-Muller
codes contain their duals, it is easy to show that 
whenever the number of rows in the stabiliser is less than $n$ (ie $k > 0$), 
$H_x \cdot H_z^T = 0$
and therefore equation (\ref{selfdual}) is satisfied
(illustrative examples will
be given shortly). When the stabiliser or generator matrix
has $n$ rows, on the other hand, one finds $D_x \cdot D_z^T \ne 0$
which, combined with the fact that the rest of
$H_x \cdot H_z^T$ is zero, means that $H_x \cdot H_z^T$ cannot be
symetric and therefore
(\ref{selfdual}) is not satisfied. In summary, the method works for
$k > 0$, and the generators for $k < 0$ are stabilisers of $k>0$ codes.

Some examples will clarify the construction.

First take the simplest case $r=2,\;t=1$. This produces 
the following generator
  \beq
{\cal G} = \left( \begin{array}{c|c}
1111 & 0000 \\
0000 & 1111 \\
0011 & 0101 \\
0101 & 0110   \end{array} \right)
  \eeq
which does not satisfy (\ref{selfdual}), but is significant in that
it `seeds' the rest of the construction method.

Next, with $r=3,\;t=1$
we obtain the $[[8,3,3]]$ code first derived independently by
several authors \cite{Gott,Cald2,SQC}:
  \beq
{\cal G}_{[[8,3,3]]} = \left( \begin{array}{c|c}
11111111 & 00000000 \\
00001111 & 00000000 \\
00110011 & 00000000 \\
01010101 & 00000000 \\
00000000 & 11111111 \\
00000000 & 00001111 \\
00000000 & 00110011 \\
00000000 & 01010101 \\
00000011 & 00000101 \\
00000101 & 00010001 \\
00010001 & 00000110   \end{array} \right)
  \eeq
The whole set of codes with $t=1$ are those derived previously by Gottesman 
\cite{Gott}. The present discussion constitutes a particularly
straightforward way of constructing them. The stabiliser of such a
code is
  \beq
{\cal H}_{[[8,3,3]]} = \left( \begin{array}{c|c}
11111111 & 00000000 \\
00000000 & 11111111 \\
00001111 & 00110011 \\
00110011 & 01010101 \\
01010101 & 00111100  \end{array} \right)  \label{H8}
  \eeq

To see more clearly why the method fails for $k=0$, consider the
following stabiliser for $t=2,\; r=4$:
   \beq
{\cal H}_{t=2,r=4} = \left( \begin{array}{c|c}
1111111111111111 & 0000000000000000 \\
0000000011111111 & 0000000000000000 \\
0000111100001111 & 0000000000000000 \\
0011001100110011 & 0000000000000000 \\
0101010101010101 & 0000000000000000 \\
0000000000000000 & 1111111111111111 \\
0000000000000000 & 0000000011111111 \\
0000000000000000 & 0000111100001111 \\
0000000000000000 & 0011001100110011 \\
0000000000000000 & 0101010101010101 \\
0000000000001111 & 0000000000110011 \\
0000000000110011 & 0000000001010101 \\
0000000001010101 & 0000001100000011 \\
0000001100000011 & 0000010100000101 \\
0000010100000101 & 0001000100010001 \\
0001000100010001 & 0000000000111100
  \end{array} \right)
  \eeq
Comparing with equation (\ref{H8}), one sees that the problem is that the
ones are too spaced out in the final 3 rows of $H_x$ for this part
to satisfy the parity checks of the previous 3 rows.
This happens whenever the relevent classical Reed-Muller code is too
small to contain its dual, which is the case for $k \le 0$ in our
construction.

Next, the stabiliser for $t=2,\;r=5$ will be shown. It is close
to optimal (see table in \cite{CaldGF4}), and is the simplest
previously unknown code obtained in this paper:
  \beq
{\cal H}_{[[32,10,6]]} =    \nonumber
  \eeq
  \beq
\left( \begin{array}{c|c}
G[32,16,8] & {\bf 0}    \\
{\bf 0}    & G[32,16,8] \\
00000000000000000000000011111111 & 00000000000000000000111100001111 \\
00000000000000000000111100001111 & 00000000000000000011001100110011 \\
00000000000000000011001100110011 & 00000000000000000101010101010101 \\
00000000000000000101010101010101 & 00000000000011110000000000001111 \\
00000000000011110000000000001111 & 00000000001100110000000000110011 \\
00000000001100110000000000110011 & 00000000010101010000000001010101 \\
00000000010101010000000001010101 & 00000011000000110000001100000011 \\
00000011000000110000001100000011 & 00000101000001010000010100000101 \\
00000101000001010000010100000101 & 00010001000100010001000100010001 \\
00010001000100010001000100010001 & 00000000000000000000111111110000
\end{array}  \right)        \nonumber
  \eeq

Before concluding, let us note that many other good quantum codes can be 
obtained either directly or indirectly by the construction shown in equation 
(\ref{GD}). I have thus been able to obtain many of the codes tabulated by
Calderbank {\em et. al.} \cite{CaldGF4}. An example of
a code obtained indirectly is the $[[5,1,3]]$ perfect code, which is
obtained by deleting any bit from the following $[[6,0,4]]$ code
constructed using equation (\ref{GD}):
  \beq 
{\cal G}_{[[6,0,4]]} = 
\left( \begin{array}{c|c}
001111 & 000000 \\
110011 & 000000 \\
000000 & 001111 \\
000000 & 110011 \\
111111 & 010101 \\
010101 & 100101   \end{array} \right)
  \eeq

I will conclude with a few comments on the usefulness of the
quantum Reed-Muller codes. Like 
their classical counterparts, these codes are easily 
encoded and decoded. This may be especially important for
applications in a quantum computer where efficient correction is
neccessary.
The codes are optimal (have maximal $k$ for given $n$ and $d$) 
for $t=1$ \cite{Gott}, and close to optimal for small $t$ and $r$
(comparing with the table of codes in \cite{CaldGF4}, it is seen that
$[[32,10,6]]$ is close to optimal).
They are far from optimal for large $t$ and $r$, but should form a good 
starting point for deriving other codes. The code size falls to zero
at $t=r/2$ 
with even $r$, giving parameters $[[2^r,0,2^{r/2} + 2^{r/2 - 1}]]$,
ie $d \simeq n^{1/2}$ for $n \gg 1$, 
which means that in the limit of large
$r$ the small size codes are no better than those obtained by
the simpler method of equation (\ref{G1G2}) applied
to the self-dual Reed-Muller codes, leading to parameters
$[[2^{r-1},0,2^{r/2}]]$. 

I would like to acknowledge helpful conversations with Daniel Gottesman
and Markus Grassl at the recent ISI Foundation workshop,
``Quantum Computation.'' The author is supported by the Royal Society.

\newpage

\begin{tabular}{lr|rrrrr}
k&     & \multicolumn{5}{c}{d}   \\
 &     &    3 &    6 &  12 &  24 & 48 \\
\hline
 &   4 &    0 &     &     &      & \\
 &   8 &    3 &  -3 &     &      & \\
 &  16 &   10 &   0 & -10 &      & \\
n&  32 &   25 &  10 & -10 & -25  & \\
 &  64 &   56 &  35 &   0 & -35  &  -56 \\
 & 128 &  119 &  91 &  35 & -35  &  -91 \\
 & 256 &  246 & 210 & 126 &   0  & -126 \\
 & 512 &  501 & 456 & 336 & 126  & -126 \\
 &1024 & 1012 & 957 & 792 & 462  &    0
\end{tabular}

Table 1. Parameters $[[n,k,d]]$ of the quantum Reed-Muller codes,
given by equation (\ref{par}), for $2 \le r \le 10$, $1 \le t \le 5$.

\end{document}